# Strongly Correlated $s$-Wave Superconductivity in the $N$-Type Infinite-Layer Cuprate

C.-T. Chen,[1] P. Seneor,[1] N.-C. Yeh,[1] R. P. Vasquez,[2] L. D. Bell,[2] C. U. Jung,[3] J. Y. Kim,[3]
Min-Seok Park,[3] Heon-Jung Kim,[3] and Sung-Ik Lee[3]

[1]*Department of Physics, California Institute of Technology, Pasadena, California 91125*
[2]*Jet Propulsion Laboratory, California Institute of Technology, Pasadena, California 91109*
[3]*National Creative Research Initiative Center for Superconductivity and Department of Physics,
Pohang University of Science and Technology, Pohang 790-784, Korea*

Quasiparticle tunneling spectra of the electron-doped ($n$-type) infinite-layer cuprate $Sr_{0.9}La_{0.1}CuO_2$ reveal characteristics that counter a number of common phenomena in the hole-doped ($p$-type) cuprates. The optimally doped $Sr_{0.9}La_{0.1}CuO_2$ with $T_c = 43$ K exhibits a momentum-independent superconducting gap $\Delta = 13.0 \pm 1.0$ meV that substantially exceeds the BCS value, and the spectral characteristics indicate insignificant quasiparticle damping by spin fluctuations and the absence of pseudogap. The response to quantum impurities in the Cu sites also differs fundamentally from that of the $p$-type cuprates with $d_{x^2-y^2}$-wave pairing symmetry.

PACS numbers: 74.72.–h, 74.50.+r, 74.62.Dh

The predominantly $d_{x^2-y^2}$ pairing symmetry [1,2], the existence of spin fluctuations in the $CuO_2$ planes [3,4], and the pseudogap phenomena [3–5] in the underdoped and optimally $p$-type cuprates have been widely conceived as essential to high-temperature superconductivity. However, recent scanning tunneling spectroscopic studies have shown that the pairing symmetry may be dependent on the hole-doping concentration, with $(d_{x^2-y^2} + s)$ mixed symmetries in certain overdoped cuprates such as $(Y_{1-x}Ca_x)Ba_2Cu_3O_{7-\delta}$ [6]. Furthermore, whether the pairing symmetry is $d_{x^2-y^2}$ or $s$ wave in the one-layer $n$-type cuprates such as $Nd_{1.85}Ce_{0.15}CuO_{4-\delta}$ and $Pr_{1.85}Ce_{0.15}CuO_{4-\delta}$ remains controversial [7,8], and it has been suggested that the pairing symmetry in the one-layer $n$-type cuprates may change from $d_{x^2-y^2}$ to $s$, depending on the electron doping level [9]. The nonuniversal pairing symmetries in cuprate superconductors imply that the symmetry is likely the result of competing orders rather than a sufficient condition for pairing. Nonetheless, an important consequence of either $d_{x^2-y^2}$ or $(d_{x^2-y^2} + s)$-wave pairing is that the resulting nodal quasiparticles can interact strongly with the quantum impurities in the $CuO_2$ planes [10,11], such that a small concentration of impurities can give rise to strong suppression of superconductivity and modification of the collective $Cu^{2+}$ spin excitations [6,12–17]. In addition, Kondo effects could be induced by nonmagnetic impurities through breaking the nearest-neighbor antiferromagnetic $Cu^{2+}$-$Cu^{2+}$ interaction [18]. Such strong response to nonmagnetic impurities is in sharp contrast to conventional $s$-wave superconductivity [19,20].

Despite significant progress in the studies of cuprate superconductivity, the research on the simplest form of cuprates, the infinite-layer system $Sr_{1-x}L_xCuO_2$ ($L$ = La, Gd, Sm), has been limited [21–23] due to the difficulties in making single-phase samples with complete superconducting volume. Recently, Jung et al. [24] have demonstrated single-phase samples of $Sr_{0.9}La_{0.1}CuO_2$ with nearly 100% superconducting volume and a sharp superconducting transition temperature at $T_c = 43$ K, thus enabling reliable spectroscopic studies of the pairing symmetry and the effects of quantum impurities. These single-phased infinite-layer cuprates are $n$-type with $P4/mmm$ symmetry, which differ significantly from other cuprates in that no excess charge reservoir block exists between consecutive $CuO_2$ planes except a single layer of Sr(La), as illustrated in Fig. 1(a), suggesting stronger $CuO_2$ interplanar coupling. Furthermore, the $c$-axis superconducting coherence length ($\xi_c = 0.53$ nm) is found to be longer than the $c$-axis lattice constant ($c_0 = 0.347$ nm) [25], in stark contrast to other cuprate superconductors with $\xi_c \ll c_0$. Hence, the superconducting properties of the infinite-layer system are expected to be more three-dimensional, as opposed to the quasi-two-dimensional nature of all other cuprates. In this Letter, we report experimental findings based on the scanning tunneling spectroscopy studies of pure infinite-layer samples and those with a small concentration (1%) of either magnetic or nonmagnetic quantum impurities. A number of surprising results are found and compared with the established phenomena in other cuprates.

The samples studied in this work included high-density granular materials of $Sr_{0.9}La_{0.1}CuO_2$ (SLCO), $Sr_{0.9}La_{0.1}(Cu_{0.99}Zn_{0.01})O_2$ (1% Zn-SLCO), and $Sr_{0.9}La_{0.1}(Cu_{0.99}Ni_{0.01})O_2$ (1% Ni-SLCO) [24]. X-ray diffraction (XRD) confirmed the single-phase nature of all samples, and both XRD and scanning electron microscopy [24] revealed random grain orientation and a typical grain size of a few micrometers in diameter. Magnetization studies revealed nearly 100% superconducting volume for all samples, with $T_c = 43$ K for SLCO and 1% Zn-SLCO, and $T_c = 32$ K for 1% Ni-SLCO. Structurally, the infinite-layer system with up to <3% Zn or Ni substitutions was stoichiometrically homogeneous [24]. However, the superconductivity appeared to be sensitive to the type of impurities. While nonmagnetic Zn had little effect on $T_c$ for up to 3% concentration, strong suppression of $T_c$ already occurred with 1% Ni, and nearly complete suppression of $T_c$ was reached with only 2% Ni [24]. Thus, the global response of SLCO to impurities appeared different from that in the $p$-type cuprates [6,12–17] and was similar to that in conventional superconductors [19,20].

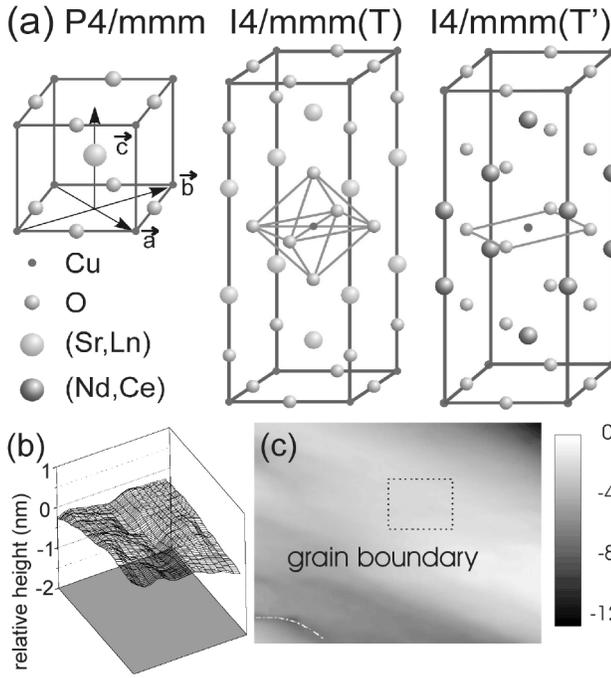

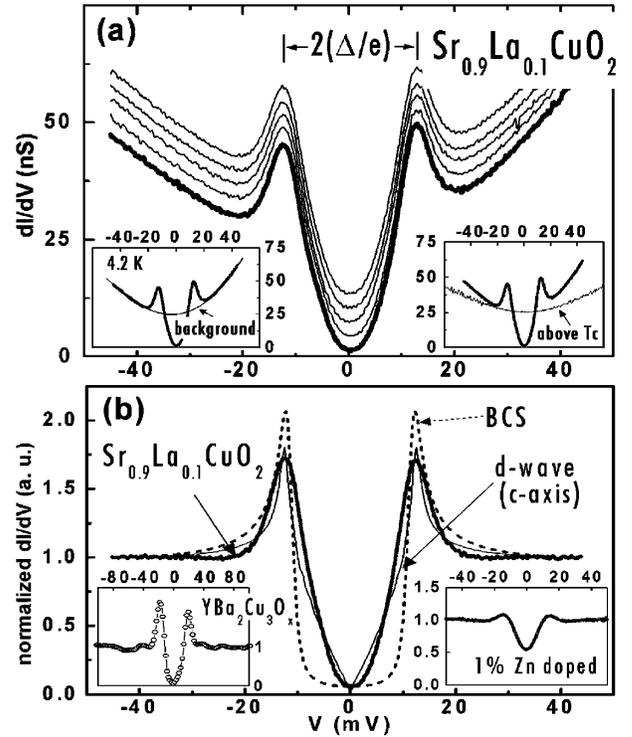

FIG. 1. (a) Comparison of the structure of the infinite-layer system $Sr_{1-x}L_xCuO_2$ ($L$ = La, Gd, Sm), with those of the one-layer $p$-type ($T$-phase) and one-layer $n$-type ($T'$-phase) cuprates. (b) A representative surface topography of an area of SLCO with subnanometer flatness. The typical area with atomic-scale flatness where most tunneling spectra were taken was greater than (20 nm $\times$ 20 nm), and the work function of the spectra was 0.1 $\sim$ 1 eV. (c) A zoom-out view of the region shown in part (b) (indicated by the dashed box) over an area (49 nm $\times$ 40 nm). Also shown in the lower left corner is a grain boundary.

Quasiparticle tunneling spectra were taken using a low-temperature scanning tunneling microscope on hundreds of randomly oriented grains for the three different infinite-layer samples, so that a range of different quasiparticle momenta relative to the crystalline axes of the local grains could be sampled. The sample surface was prepared according to the chemical etching method described elsewhere [26], and a nearly stoichiometric surface with no discernible chemical residue was confirmed with the x-ray photoemission spectroscopy [26]. A typical surface topography of the pure SLCO sample for our spectroscopic studies with subnanometer flatness is exemplified in the left panel of Fig. 1(b), and a zoom-out view of this area is illustrated in Fig. 1(c). Confirming the local flatness for the tunneling spectra was to ensure that the average momentum of the incident quasiparticles relative to the crystalline axes of a grain was well defined. A set of representative differential conductance ($dI/dV$) vs biased voltage ($V$) spectra for such a flat area is given in Fig. 2(a). In general, all spectral characteristics revealed long-range ($>$50 nm) spatial homogeneity within each grain and small variations in the superconducting gap value ($\Delta = 13.0 \pm 1.0$ meV) from one grain to another. Here ($2\Delta/e$) was defined as the conductance peak-to-peak separation in the spectra. This observation was in sharp contrast to our previous findings of strongly momentum-dependent spectra in the $p$-type cuprates with $d_{x^2-y^2}$ pairing symmetry [6]. The absence of the zero bias conductance peak (ZBCP) [6], known as a hallmark for unconventional pairing symmetry, for over 1000 spectra provided additional support for a fully gapped Fermi surface.

FIG. 2. (a) Representative $dI/dV$ vs $V$ quasiparticle spectra of SLCO taken at 4.2 K. The curves correspond to spectra taken at $\sim$1.5 nm equally spaced locations within one grain and have been displaced vertically for clarity except the lowest curve. Left inset: a typical spectrum taken at 4.2 K (solid line) compared with the corresponding high-voltage background (dashed line). Right inset: comparison of a typical spectrum taken at 4.2 K with one taken slightly above $T_c$. (b) A spectrum normalized relative to the high-voltage background given in the left inset of (a), together with a BCS theoretical curve for the normalized DOS at $(T/T_c) = 0.1$ and a corresponding $c$-axis tunneling spectrum for a pure $d_{x^2-y^2}$-wave superconductor (thin solid line). Left inset: a normalized $c$-axis tunneling spectrum of an optimally doped $YBa_2Cu_3O_{7-\delta}$ ($T_c = 92.5 \pm 0.5$ K). Right inset: a typical spectrum for the 1% Zn-SLCO sample taken at 4.2 K.

Despite suggestive evidence for $s$-wave pairing symmetry, the unusually large ratio of $(2\Delta/k_BT_c) \approx 7.0$ as compared with the BCS ratio of 3.5 was indicative of strong coupling effects. Moreover, the commonly observed "satellite features" in the quasiparticle spectra of $p$-type cuprate superconductors [6], as exemplified in the left inset of Fig. 2(b), were invisible in SLCO. The satellite features in $p$-type cuprates were associated with quasiparticle damping by many-body interactions such as the collective spin excitations [6,27,28]. Thus, the absence of satellite features in SLCO is consistent with weakened spin fluctuations as the result of diluted antiferromagnetic coupling due to the presence of $Cu^{1+}$ introduced by electron doping. In addition, $\Delta$ was found to completely vanish above $T_c$, with no apparent energy scale associated with any depression of the density of states (DOS) at $T > T_c$, as shown in the right inset of Fig. 2(a), and the tunneling spectra were nearly temperature independent from just above $T_c$ to $\sim 110$ K. The absence of any spectroscopic pseudogap in the $n$-type infinite-layer system was distinctly different from the findings in optimally doped and underdoped $p$-type cuprates [5] and was independently verified by the NMR studies on similar samples [29].

By normalizing a typical spectrum in Fig. 2(a) relative to the background conductance shown in the left inset of Fig. 2(a), we compared the quasiparticle DOS of SLCO with the BCS theoretical curve, as illustrated in Fig. 2(b). The spectral weight of SLCO for quasiparticle energies at $|E| \geq \Delta$ was smaller than the BCS prediction, whereas additional DOS appeared for $|E| < \Delta$ and the DOS approached 0 at the Fermi level (i.e., $V = 0$). Such behavior cannot be accounted for by the simple inclusion of disorder in the BCS weak-coupling limit, because the latter would have only broadened the width of the conductance peaks and also increased the DOS near $V = 0$ substantially. The spectra also differed fundamentally from those of pure $d_{x^2-y^2}$-wave cuprates [6] because of the lack of discernible gap variations and of the absence of ZBCP in all spectra taken on random grain orientations. Even in a special case of $c$-axis tunneling, $|d^2I/dV^2|_{V\to 0^\pm}$ would have been a positive constant in a $d_{x^2-y^2}$-wave superconductor, as simulated by the thin solid line in Fig. 2(b), which is in contrast to the finding of $|d^2I/dV^2|_{V\to 0^\pm} = 0$ in SLCO. Interestingly, recent Knight shift data from NMR studies of similar SLCO samples have revealed much smaller normal-state DOS at the Fermi level as compared with those of other cuprates [29], which corroborates the inapplicability of weak-coupling theory to SLCO. We therefore suggest strongly correlated $s$-wave pairing in the infinite-layer system based on the empirical findings of momentum-independent quasiparticle spectra, absence of ZBCP, and $|d^2I/dV^2|_{V\to 0} = 0$ for all grain orientations.

In the case of 1% Zn-SLCO, the spectral characteristics also revealed long-range spatial homogeneity in the spectra and a similar gap value ($\Delta = 13.0 \pm 2.5$ meV) for randomly sampled areas in different grains, as exemplified in the right inset of Fig. 2(b). Given that the average separation among Zn impurities is $\sim (1.8 \times 1.8 \times 1.6)$ nm$^3$, our exhaustive spectral studies should have covered a significant number of Zn impurities. However, no significant local variations were found in the spectra of the 1% Zn-SLCO, which differed fundamentally from our observation of atomic-scale spectral variations in a $YBa_2(Cu_{0.9934}Zn_{0.0026}Mg_{0.004})_3O_{6.9}$ single crystal near nonmagnetic Zn or Mg impurities using the same apparatus [6]. Nevertheless, the conductance peaks in 1% Zn-SLCO were significantly broadened relative to pure SLCO, with an increase in the DOS for $|E| < \Delta$, as shown in the right inset of Fig. 2(b). These features suggest that Zn impurities resulted in reduced quasiparticle lifetime while retaining $T_c$, similar to the response of conventional $s$-wave superconductors [19,20].

In contrast, two types of spectra were observed in 1% Ni-SLCO, as illustrated in Fig. 3(a). The majority spectra (>90%) exhibited suppressed coherence peaks, large zero bias residual conductance, strong electron-hole spectral asymmetry, and gradual spatial evolution over a long range. In contrast, the minority spectra (<10%) exhibited sharp spectral peaks, small zero bias conductance, and varying electron-hole spectral asymmetry over a short range (<1 nm), as exemplified in the inset of Fig. 3(a) for two representative minority spectra. The significant spectral asymmetry implied different phase shifts in the electronlike and holelike quasiparticle states as the result of broken time-reversal symmetry [20,30], which may be responsible for the global suppression of the superconducting phase coherence and thus a reduction in $T_c$.

Assuming homogeneous Ni-impurity distributions, the average Ni-Ni separation would be $d_{Ni} \sim 1.8$ nm in the $ab$ plane and $\sim 1.6$ nm along the $c$ axis in each grain. The impurity wave function with poor screening from the carriers would have extended over a coherence volume ($\xi_{ab}^2 \xi_c$) [20,30]. Given the coherence lengths $\xi_{ab} \sim 4.8$ nm and $\xi_c \sim 0.53$ nm [25], $\sim 30\%$ volume probability in each grain could be considered as under significantly weaker impurity influence. In the limit of completely random grain orientation in 1% Ni-SLCO, the STM studies of the grain surfaces would have $\sim 20\%$ probability for finding surface regions with weak impurity influence and spatial extension over a short range ($\sim 0.5$ nm) along the $c$ axis. This simple estimate is in reasonable agreement with our observation of ∼10% minority spectra with short-range (<1 nm) spatial homogeneity. However, due to the lack of direct information for the Ni distribution on the sample surface, the true origin for two types of spectra in 1% Ni-SLCO remains uncertain.

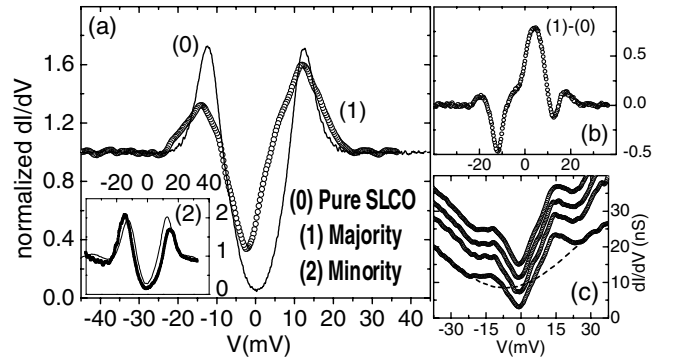

FIG. 3. (a) Main panel: comparison of a normalized majority spectrum of 1% Ni-SLCO and that of pure SLCO at 4.2 K. The normalization was made relative to the background conductance shown by the dashed line in part (c). Inset: two minority spectra with different electron-hole asymmetry. (b) Spectral difference of the majority spectra relative to that of the pure SLCO. (c) A series of spectra taken on the same grain of 1% Ni-SLCO at $\sim 3$ nm apart. The conductance of all curves except the lowest one has been displaced up for clarity.

Considering the spectral difference between the majority spectrum of 1% Ni-SLCO and that of pure SLCO, as shown in Fig. 3(b), we find that the spectral characteristics resemble the findings in Ref. [20] and are representative of the impurity-induced state. On the other hand, the slowly varying majority spectra of 1% Ni-SLCO, as shown in Fig. 3(c), were possibly the result of strong overlapping in the Ni-impurity wave functions and of weak screening effects due to low carrier density in SLCO, which differed markedly from the rapidly diminishing impurity effects away from an isolated Mn or Gd atom on the surface of Nb [20], and also from the strong atomic-scale spectral variations near Ni impurities in the $p$-type cuprate $Bi_2Sr_2Ca(Cu_{1-x}Ni_x)_2O_{8+x}$ [17]. The contrast in the spatial extension of the Ni-impurity effects may be attributed to the variation in the impurity coupling strength and range, and also to the degree of impurity screening by carriers. We suggest that the strongly interacting Ni impurities in 1% Ni-SLCO are analogous to a Kondo alloy, which cannot be explained by the Abrikosov-Gor'kov theory for magnetic impurities in BCS superconductors [30].

The parent materials of all $p$-type cuprates are Mott insulators with strong on-site Coulomb repulsion [3,4]. Thus, the formation of $d_{x^2-y^2}$-wave pairing symmetry is energetically favorable in reducing the Coulomb repulsion while retaining the quasi-two-dimensionality. On the other hand, the strong three-dimensional coupling in the infinite-layer system could favor $s$-wave pairing symmetry by compensating the resulting increase in the Coulomb repulsion with a large gain in the condensation energy. Thus, the pairing symmetry of cuprate superconductors may be dependent on the specific structures and various competing energy scales. Similarly, the pseudogap phenomena may be due to competing orders and need not be universal for all cuprates.

A recent study of the angular-resolved photoemission spectroscopy (ARPES) on three different families of $p$-type cuprates suggested that an abrupt change of the electron velocity in the 50−80 meV energy range was ubiquitous and might be associated with the longitudinal optical oxygen phonon modes in the $CuO_2$ planes [31]. Such changes in ARPES approximately coincided with a "dip" feature in the quasiparticle tunneling spectra of some cuprates [6]. However, our tunneling spectra of SLCO revealed a dip energy at ∼20 meV, and that of $YBa_2(Cu_{0.9934}Zn_{0.0026}Mg_{0.004})_3O_{6.9}$ at ∼30 meV, much smaller than the energy ∼50 meV for pure $YBa_2Cu_3O_{6.9}$ [6]. Thus, the only ubiquitous features among all cuprates appear to be the strong electronic correlation and the background antiferromagnetism of $Cu^{2+}$ ions in the $CuO_2$ planes.

The research at Caltech was supported by NSF Grant No. DMR-0103045 and the Caltech President's Fund. Part of the work was performed by the Center for Space Microelectronics Technology, Jet Propulsion Laboratory, and was sponsored by NASA. The work at Pohang University was supported by the Ministry of Science and Technology of Korea through the Creative Research Initiative Program.